%% file: articoloRivista-SUBhighSpeed.tex
\documentclass{article}

\usepackage{amsmath}
\usepackage[a4paper]{geometry}
\usepackage{graphicx}
\usepackage{amsfonts,subfigure,multirow}
\usepackage{microtype}
\usepackage{siunitx}
\usepackage{booktabs}
\usepackage[colorlinks=false, pdfborder={0 0 0}]{hyperref}
\usepackage{cleveref}

\usepackage{setspace}
\usepackage[printfigures]{figcaps}

\usepackage{etoolbox}
\makeatletter
\patchcmd{\@figurepage}{\vspace{20pt}}{\clearpage}{}{}
\patchcmd{\@tablepage}{\bigskip}{\clearpage}{}{}
\makeatother

\begin{document}
\title{Performance issues in content dissemination to metropolitan mobile users}

\author{Giuseppe Corrente\\
Dipartimento di Informatica Universit\`a di Torino\\
Corso Svizzera 185 - 10149 Torino, Italy\\
Tel.: +39-011-6706711\\
Fax: +39- 011-751603\\
Email: corrente@di.unito.it
}

\maketitle

\begin{abstract}

In this paper we consider a set of heterogeneous terminals
in a urban area that are interested in collecting the information originated from
several sources. This set includes
mobile nodes (pedestrian and vehicles) and fixed terminals.
In particular, each terminal aims at retrieving the data items in a limited
region of interest (ROI) centered around the node position.
Since data items may change over time all nodes must strive
for having access to the latest version.

The goal of the paper is to evaluate the amount of information
each node is able to gather (coverage) resorting to simple
distributed data collection and sharing through local broadcast
communications.

We study the diffusion of information updates in the whole area,
evaluate the impact of energy saving policies in the protocol
version run by pedestrian devices, and the impact of contextual
awareness about location and motion of nodes in the forwarding policies.

The study we present in this paper has been carried out
through simulation.
To this end we develop a discrete event simulator working on top of mobility
and radio propagation traces obtained from the
UDelModels tools that allow to  obtain realistic traces of mobility and radio propagation.

\end{abstract}



\pagebreak





\openup 1em

\input{new_INTRO-art.tex}

\input{related.tex}

\input{SYSTEM.tex}
\input{METHODOLOGY.tex}

\input{results.tex}

\input{CONCLUSION.tex}

\section*{Acknowledgement}

I would like to thank Professors Gaeta Rossano, Grangetto Marco and Sereno Matteo for their expert advice
and all the Computer Science Department of the University of Turin for their invaluable support.


\bibliographystyle{abbrv}
\bibliography{biblioUrb}

\end{document}

%% file: new_INTRO-art.tex
\section{Introduction}
\label{INTRO}
The development of sophisticated distributed applications for spreading informational and environmental data to metropolitan mobile users
has become more feasible thanks to 
the widespread use of battery-operated smart terminals equipped with powerful multi-core processors, large random access memories, and wireless communication capabilities. Data collection and distribution lie at the heart of such applications: these two abstractions are implemented using protocols whose performance are influenced by numerous factors. To name a few:
\begin{itemize}
\item physical characteristics of the environment where the application is run: the ability of a terminal to communicate data to others within radio range is dependent on the presence of obstacles, e.g., terminals within buildings in a urban scenario may not receive data transmitted by terminals moving along the adjacent streets outside or residing on different floors of the same building.
\item terminals heterogeneity: terminals with different transmission powers (hence different radio ranges) can communicate with a number of other terminals that depend on it. Or, terminals moving at different speeds and following different mobility patterns can count on a different neighborhood for communication.
\item information dynamics: if data to be disseminated is time-dependent then special storage and forwarding policies have to be designed.
\end{itemize}
To make the design of these applications even more complex, performance must be traded-off against energy consumption since mobile terminals are battery-operated. This very short and incomplete list of factors  shows that performance evaluation of protocols to support distributed applications for content dissemination to metropolitan mobile users in realistic scenarios is absolutely essential.

\subsection*{Our contribution}

In this paper we analyze the performance of a simple data exchange protocol among mobile nodes (operated by pedestrians and vehicles) and fixed terminals in a urban scenario. The area under investigation is ideally partitioned into a grid where each subregion is the source of one data item (e.g., autonomous sensing of environmental signals, reading from active sensors, passive information conveyed by RFID tags, commercial advertisements or traffic data).
Nodes are placed in the environment and aim at retrieving the data items in a limited {\em region of interest} (ROI). A ROI can be centered around the current node position or detached from it and fixed for a given amount of time. In the latter case, the association between nodes and ROIs can be one to one or many to one, i.e. some nodes may share the same ROI. Clearly,  for mobile nodes ROI is a time varying concept due to the dynamic behavior of pedestrians and vehicles. Since data items may change over time all nodes must strive for having access to the latest version; we characterized the protocol performance by an index we call \emph{coverage} that is defined as the fraction of updated data items collected by a node with respect to the total amount of accessible data items within its ROI at a specific time.

Data collection is possible through acquisition and local broadcast of data items. Each mobile node can acquire the data item in a specific location as soon as its current position falls within the subregion. Nodes also transmit (broadcast to nearby nodes) data items in a range that depends on their radio coverage. Nodes are able to store data items but their memory size depends on the type of terminal: hand-held computing devices operated by pedestrians are assumed to reserve a small memory to the data collection application while terminals mounted on fixed terminals and vehicles are assumed to use a larger memory. Furthermore, mobile nodes whose computing device is battery operated (pedestrians are assumed to use this kind of terminals) are required to cautiously transmit their data items in order to preserve energy and prolong the device operating time.

We study the diffusion of information updates in the whole area, 
and evaluate the impact of energy saving policies in the protocol 
version run by pedestrian devices. In addition we estimate the impact of contextual 
awareness about location and motion of nodes in the forwarding policies.
One of the results we find is that
most of the transmissions of these terminals are redundant and do not help in increasing the coverage of ROIs of neighbor nodes. Therefore we consider the performance of nodes when nodes operated by pedestrians transmit at a much lower rate. We show that huge saving in overall number of transmissions (and hence in energy consumption) can be achieved at the cost of a reasonable reduction of the coverage of all nodes types. We also show that users within buildings may have poor performance due to difficult spreading 
of information inside a building because their of limited radio coverage. 
We evaluate the impact of of the introduction of 
a very simple form of communication infrastructure represented 
by relaying nodes placed in the elevators of buildings.

All results we present are obtained from a simulator developed in C++ on top of the UDelModels tools \cite{mobSim,propSim}. These tools are used to define three-dimensional maps of urban areas and to obtain traces of mobility and radio propagation. The mobility characteristics of the nodes are based on statistical studies of population and traffic dynamics. Mobility traces generated by UDelModels are very detailed: for instance, pedestrians exhibit a motion that is representative of people in an urban scenario, with different mobility distributions for outdoor and indoor walking, respectively. Moreover, a typical daily human activity cycle is taken into account. Cars mobility patterns take into account speed limits and traffic lights. The UDelModels propagation simulator is used to estimate the point to point channel loss between each pair of nodes in the three-dimensional space, taking into account the urban three-dimensional profile. The usage of such a detailed multi-dimensional mobility and propagation model for the analysis of content dissemination policies represents the major novelty of our work, allowing us to identify critical features that may remain hidden when working with a simplified flat world with random mobility.

The paper is organized as follows: Section \ref{sec:related} describes related works, Section
\ref{sec:system} presents the system model, Section \ref{sec:methodology} describes the simulator we developed and the tools we used, and Section \ref{sec:simulator} discusses the results we obtained. Finally, Section \ref{sec:conclusions} outlines conclusions and future perspectives of the current work.

%% file: related.tex
\section{Related works}
\label{sec:related}

Efficient information diffusion in wireless ad-hoc networks is a challenging goal that has received considerable
attention in many recent studies.


Many researchers in the field of wireless sensor networks have tackled this issue.
In \cite{expl} a three tiers architecture called MULE was proposed. MULE leverages  light sensors of limited memory capacity carried by people, animals, and vehicles to buffer and forward data gathered by a set of sensors to some access points. The nodes mobility is exploited to propagate the information while saving sensor energy.
The MULE architecture encompasses sensors 
able to read environmental data and mobile nodes aiming at forwarding the sensed information
to a set of sinks represented by fixed nodes. 
The paper shows that one can leverage node mobility to achieve significant energy savings
with respect to a more classical ad hoc network. Compared to \cite{expl}, our
work presents a generalized scenario with the introduction of dynamic ROIs, as opposed
to static sink nodes collecting the information.
ZebraNet \cite{ZebraNet} and SWIM \cite{whales} are other examples of data collection
by fixed sinks by means of exploiting mobility in the framework of wild life data tracking.

Similar approaches have been extended to the wider area of delay tolerant networks (DTN).
The PROPHET \cite{probabilisticrouting} routing protocol exploits
the fact that human and vehicular mobility is not purely random:
nodes that meet frequently in the recent past are likely to do so in the next future.
PROPHET aims at improving the routing mechanism based on the knowledge of the nodes
encounters history.
PROPHET+ \cite{prophetPlus} further builds on the same concept taking into account
energy consumption and memory requirements.
A distributed technique to collect data in a given area of interest is proposed in \cite{amorph}. 
To this end the memory of the nodes is managed so as to maximize the mutual geographical distance among the stored data in order to maximize the covered area.
In \cite{amorph} each node aims at collecting all data items by querying few neighbors
 and energy saving is obtained by limiting the usage of multi hop routing.

A sophisticated scheduling strategy and a cache dropping policy based on local history statistics are used
in \cite{2015knapsack} and \cite{2016novel}.
The aim of \cite{2016novel} is to improve a mix of delivery, delay and overhead.
However, with limited size buffer, the recent sophisticated cache strategies in \cite{2016novel} 
outperform many simpler cache management strategies of no more than 35\%.

In \cite{inf_framw} an information theoretical approach is presented to optimize
the caching mechanism used by mobile devices; in this case every node is assumed to query for the information in a given point in space, assuming a certain probability distribution for the location of the queries.
In \cite{data-centric} the problem of efficiently answering real-time geo-centric ad-hoc queries is considered. In this proposal a storage node is elected dynamically (achieving load-balancing) and a point-to-point routing scheme is used to deliver the sensed items to the corresponding storage node. 
\cite{oppCont} uses occasional contacts among people, i.e. group of student in their simulations, to distribute a digital
content pushed by few fixed nodes. They show how contacts among different group of students are crucial in the
dissemination process.
In \cite{node_coop} the authors study the packet delivery rate in DTN for different routing protocol  and degrees of node cooperation and they find that also a modest level of cooperation increases network delivery rate significantly.
A similar observation is made in \cite{non-coop}. 
The work \cite{cache} aims at maximizing the hit ratio of queries issued by mobile nodes searching for content cached by other nodes. 
In this case every nodes caches content for a period of time that depends on the distance, measured in number of hops, 
between source and destination. In \cite{Palmieri2013} and \cite{Palmieri2017} percolation-driven and Bayesan resource discovery policies are studied; these are two service discovery 
policies in infrastructure-less networks supported by parallel random walk-based searchs and Bayesan decision criterion.


In \cite{hovering,floating,locus} spatial correlation between the data and the origin of the data is exploited. In \cite{oppCoverage} opportunistic coverage metric is proposed for performance evaluation of people centric sensing applications.

Theoretical bounds on gossip performance and information diffusion speed in mobile systems have been obtained in \cite{imp,bounds} and \cite{speed}, respectively.
The capacity of ad hoc wireless networks is evaluated under general mobility conditions in \cite{gare07}.
Other approaches for the modeling and exploitation of nodes mobility in wireless networks are presented in \cite{graph} and \cite{radiograph}, where random graph theory is employed.

%% file: SYSTEM.tex
\section{System description}
\label{sec:system}
%
%
In this section the characteristics of the system under analysis are presented.
We analyze the performance of data exchange protocols among mobile nodes (operated by pedestrians and vehicles) and fixed terminals in a urban area.
Each communication node is characterized by its mobility characteristics, radio coverage and amount of memory devoted to the data exchange protocol. For the sake of our analysis a node is able to communicate a message
in area covered by its radio using broadcast transmissions.
In this paper we consider the following classes of nodes: 
\begin{itemize}
\item  {\em fixed} (F) nodes, i.e.,
wireless relay nodes or access points, that are placed at road intersections and inside buildings.
\item {\em pedestrians} (P) nodes, that carry portable devices with limited power and memory capacities. Clearly, P nodes move along streets at walking speed, typically concentrating inside buildings;
\item {\em vehicular} (V) nodes that move faster along trajectories  constrained by roads; moreover, we assume that this class of nodes has no strict limitations in terms of power and memory since they represent car-installed hardware.
\end{itemize}
In our experiments we noted that the dissemination of information between different levels of a
building can be very critical, due to the limited radio propagation
between adjacent floors. To overcome this problem a new class of semi-fixed nodes has been introduced in the system. In particular, we assume to equip some elevators with a wireless relay able to exploit the vertical mobility pattern to disseminate information among different floors of a building. We call these nodes: \emph{elevator nodes} (E) nodes.

The nodes are placed in a three-dimensional space where roads and buildings with multiple levels are modeled as described in Section \ref{sec:methodology}.
A piece of information is associated with every square tiles of $\delta\times\delta$ meters. In particular we denote as $i(x,y,t)$ the information
carried by tile at location $(x,y)$ and time-stamp $t$. This abstraction is used to model the presence of pieces of information associated to the location. As an example we can think about information of social utility, e.g., traffic and mobility management, public services and events advertising, etc.
Another potential source of local information is the presence of a grid of  smart objects and/or sensors that can provide useful information associated to their position according to the Internet of Things paradigm \cite{IoTAtzori}.
The information $i(x,y,t)$ can be composed by a set of environmental measurements taken autonomously by the mobile nodes or
communicated by an infrastructure of active sensors, passive information conveyed by RFID tag, commercial advertisements
or traffic data, etc.

\begin{figure}[b]
\centering
\includegraphics[width=84mm]{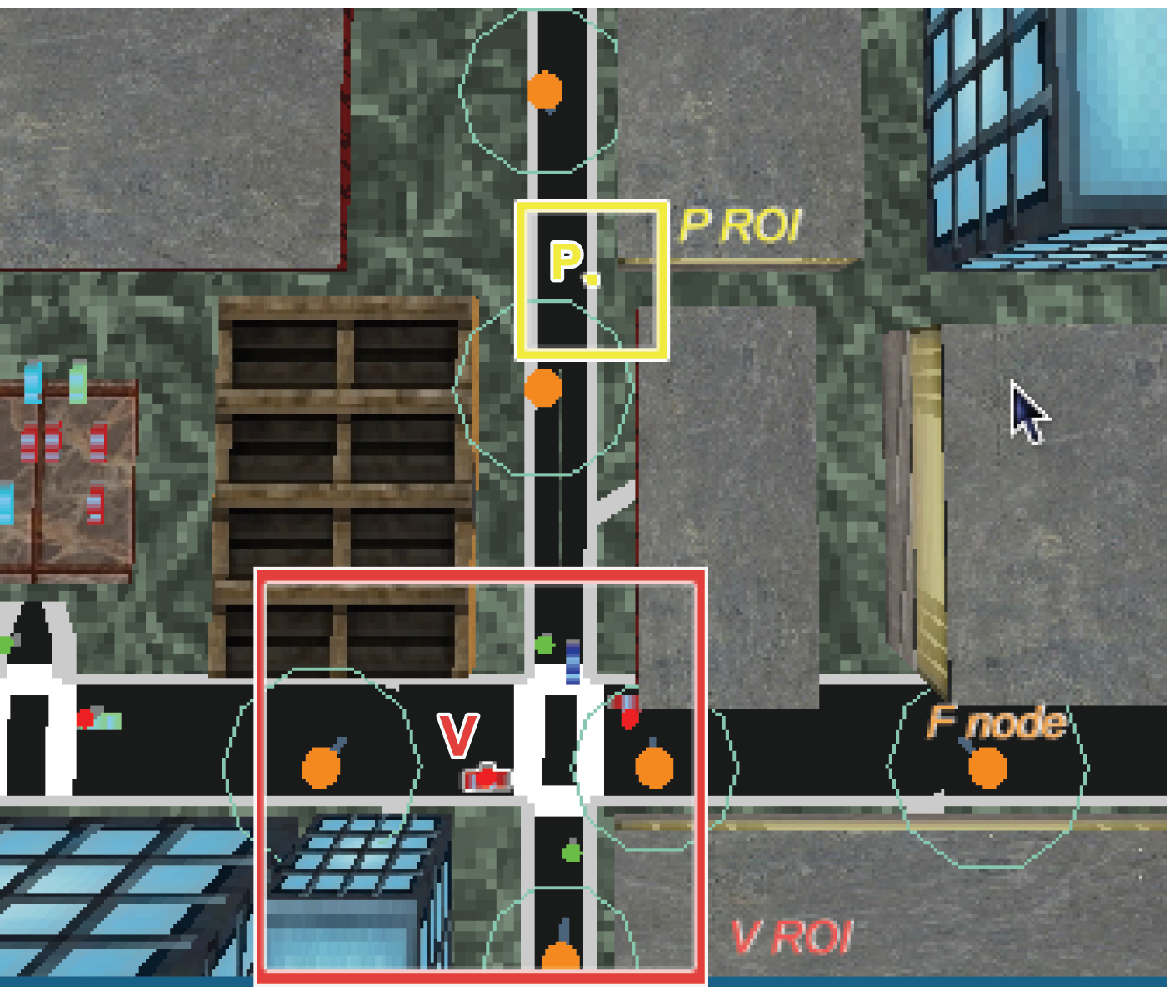}
\caption{Example of two ROIs for P and V nodes. \label{fig:scenario}}
\end{figure}

The aim of each node is to collect information associated to a given set of tiles that determine its {\em Region of Interest} (ROI). For the sake of simplicity a ROI is identified with a $\Delta\times\Delta$ square centered around a given target location.
ROIs positions may be static or dynamic in both space and time.
In Figure \ref{fig:scenario} two ROIs are shown as squares of different dimensions centered around a pedestrian and a car, respectively.

Each node aims at maximizing the knowledge of the information associated with its ROI, constrained by the size of the memory (that we denote as $B$) that has been allocated to the purpose.
To allow complete coverage of the ROI we must assume that $B \geq \left(\frac{ \Delta}{\delta} \right)^2$.

\subsection{Data exchange strategies}

Every node can complete its task, i.e. to collect the maximum amount of  information associated with the ROI, either by directly reading an interesting item $i(x,y,t)$ from the environment, or by overhearing the same piece of information from the broadcast transmissions of other nodes.
The former case occurs when a node hits an interesting location $(x,y)$ thanks to its mobility. The latter case is more interesting to model and investigate, being dependent on the mobility of all nodes in the system and the policies they adopt to collect, share, and broadcast the information stored in their memory.

In our analysis we assume that the time is slotted.
In each time slot a node acquires the information $i(x,y,t)$ corresponding to its current position (if it is not already contained in $B$). Moreover,
each node can use a simple and random broadcasting policy to propagate its own information to its neighborhood.
In particular, a node can randomly selects $k$ information items stored in its local memory $B$ and propagate them.
The nodes adopt broadcasting policies aiming at limiting power consumption and at the same time avoiding to clutter the shared radio spectrum. To this end, broadcasting is activated only once
every $T$ time slots; moreover, we investigate context aware approaches, e.g., a mobile node starts broadcasting only when it enters a new area, trying to maximize the information diffusion.

Different buffer management strategies have been considered. The basic policy is based on First In First Out (FIFO)
buffer management.
With this strategy each received piece of information  $i(x,y,t)$ is buffered with a FIFO policy if no information about the position $(x,y)$ has been included yet; on the contrary, if $i(x,y,t_0)$ is stored in the node buffer with $t_0 < t$ then, the corresponding record is updated.
The basic configuration is compared with two simple improvements, referred in the following as {\em selective dropping} (SD) and {\em selective insertion} (SI),  that aim at prioritizing the storage of those information
items that are likely to fall within the ROI.
According to SD the items can be popped out of memory $B$
only if they refer to a location outside the node's  ROI.
Similarly,  the SI prevents from storing
any data outside the node  ROI.
In \cite{PE_WASUN2011} we found that SI buffer management strategy was better than SD.

The performance index we are interested in is represented by the percentage of 
ROI covered by the items stored in the local buffer of each node.
In the following we will refer to such percentage as \emph{coverage} ($0 \le C \le 1$). The value of the coverage clearly varies from node to node and it is time dependent.
Furthermore, some subareas of the urban scenario are not accessible to nodes, e.g., due to some physical obstacle, therefore $C$ is computed as the number of items in the local buffer over the total number of \emph{accessible} elements in the node ROI.
Finally, if the environmental information changes with time only the most up-to-date elements are used for the computation of the ROI coverage.

%% file: METHODOLOGY.tex
\section{Simulation methodology}
\label{sec:methodology}

In our investigation we evaluate the impact of a realistic 3D environment where pedestrian and cars co-exist in a typical urban scenario; to this end, we adopt the precise mobility simulator UDelModels \cite{mobSim}.
UDelModels are a set of tools developed to model urban mesh networks that emulates both the mobility and the radio signal attenuation between any two nodes.
UDelModels simulate the  mobility of P and V nodes in a 3D space exploiting 
statistical studies of population and traffic dynamics. 
In UdelModels the number of nodes in the different classes and the statistical parameters of the mobility
model can be properly configured.
Pedestrians exhibit a motion that is representative of people in an urban scenario,
with different mobility distributions for outside and inside walking, respectively.
Moreover, a typical daily human activity cycle is taken into account.
Cars mobility patterns take into account speed limits and traffic lights.

The UDelModels propagation \cite{propSim} simulator is used to estimate the point to point
attenuation between each pair of nodes in the 3D space.
The channel loss is used to model the radio contacts among the communicating nodes. 
Any two nodes are assumed to be able to communicate to each other if the channel loss is below a threshold $\alpha$ (in all our experiments $\alpha$ was fixed to $-30$ dB ).

The UDelModels results are used as input to a C++ simulator we developed whose major functionalities are:
\begin{itemize}
\item to simulate the radio contacts among the nodes as far as the channel loss is below a threshold $\alpha$;
\item to implement the management of the nodes buffer;
\item  to simulate  broadcast;
\item to implement different strategies for energy saving based on transmission delays and context awareness;
\item to simulate the E nodes;
%
\item  to estimate the system performance in terms of the coverage of the nodes ROI.
\end{itemize}

\subsection{Modeling assumption}

In our simulation we do not explicitly consider collisions among concurrent transmissions of nearby nodes.  We argue that this approximation is acceptable based on the following reasoning: assume that $N$ transmitters share a wireless channel whose capacity is $M$ b/s. Each transmitter has $D$ bits to transmit in one transmission that may occur at any point in time in an interval of $T$ seconds. Transmission of $D$ bits requires $H=\frac{D}{M}$ seconds to complete. Clearly, the probability of two colliding transmissions is equal to $p_c=\frac{2H}{T}$. A more accurate expression can be obtained by considering border effects that show up  if transmissions of either nodes start within the first or the last $D$ seconds in the time interval of length $T$: in this case we obtain $p_c=\frac{6HT-4H^2}{2T^2}$. We can now derive the probability that $N$ transmitters do not collide as $p_{ok}=(1-p_c)^{\binom{N}{2}}$.
In our settings, we assume that the relevant information regarding node coordinates and time stamps (which can be extracted from GPS records) plus the size of the data items requires 30 bytes to be stored. Since 
we consider transmissions of $k=5$ data items we obtain $D=240 \cdot k$ bits to be broadcast once every $T=0.2$ seconds (see Table \ref{tab:settings}). We observed that the average number of potentially colliding nodes in all our simulations has been $N=6$. If we assume a $M=10$ Mb/s channel we obtain $p_{ok}=0.973$ which is rather high and confirms that neglecting collision phenomena represents an acceptable approximations.

%% file: results.tex
\section{Results}
\label{sec:simulator}

In this section we evaluate the performance of content dissemination to metropolitan mobile users under several scenarios. First of all Section \ref{sec-settings} describes the system settings that are common to all experiments. In Section \ref{sec-information-policies} we 
evaluate the impact of policies aimed at reducing the overall number of transmissions for energy-saving purposes when information is dynamic, in Section \ref{sec-elevators} we analyze how performance is improved for indoor users when elevators are equipped with terminals able to relay data items. Finally, in Section \ref{sec:roi} we analyze coverage for several models of ROI.

\subsection{System settings}
\label{sec-settings}

We consider a metropolitan area whose size is 550 m $\times$ 500 m, representing
9 blocks of Chicago\footnote{The map data are available at {\textit http://udelmodels.eecis.udel.edu/}}. The 3D world is partitioned into flat square tiles whose side is $\delta=25$ m and each tile is associated to a piece of information. The size of the ROI depends on the node class, with $\Delta=400,200$ and $100$ m for F, V and P nodes, respectively. In practice we assume that V nodes
are interested in a larger area with respect to P nodes because of their larger speed. F nodes have the largest ROI since we assume that there are no strict constraints on the memory of the fixed infrastructure.

The memory constraint of each node is equal to the overall number of pieces of information in the node ROI, i.e. $B=\left( \frac{\Delta}{\delta} \right)^2$.   Following the results obtained in \cite{PE_WASUN2011} the memory policy is SI/SD for both information sensing and communication. The duration of one time slot is equal to 0.2 second and each node transmits once every $T=0.2$ seconds. Each  transmission involves the fully random selection of $k=5$ data items in its buffer. 

The fixed wireless infrastructure comprises $N_F=54$ F nodes placed at road intersections and building entrances; the population we consider comprises $N_P=200$ P nodes and $N_V=50$ V nodes. The mobility of V and P nodes is simulated with UDelModels with the speed of V nodes in the range $(25, 67)$ km/h whereas the speed of P nodes is in the range $(2.5, 6.5)$ km/h. 
Furthemore, the number of people and vehicles effectively present in the simulated area at different times of the simulation is determined by UDelModels.
The channel loss is used to model the radio contacts among the communicating nodes; any two nodes are assumed to be able to communicate with each other if the channel loss is below a threshold $\alpha$ that in all our experiments was fixed to $-30$ dB.

The simulations are worked out in the time interval between 7 and 10 a.m. and the initial $15$ minutes are always considered as transient and therefore excluded from the computation of the average coverage $C$. All the common system settings are summarized in Table \ref{tab:settings}.

\begin{table}[tb]
\centering

\caption{System settings.}
\label{tab:settings}
\begin{tabular}{|c|c|c|c|}
\hline
Parameter & \multicolumn{3}{|c|}{Value} \\
\hline
Simulated area & \multicolumn{3}{|c|}{550 m $\times$ 500 m } \\
$\delta$ & \multicolumn{3}{|c|}{25 m } \\
$B$ &  \multicolumn{3}{|c|}{$\left( \frac{\Delta}{\delta} \right)^2$} \\
$k$  & \multicolumn{3}{|c|}{5} \\
$T$  & \multicolumn{3}{|c|}{0.2 s} \\
$\alpha$  & \multicolumn{3}{|c|}{-30 dB}  \\
Simulated time & \multicolumn{3}{|c|}{from 7 to 10 a.m}\\
Confidence interval  & \multicolumn{3}{|c|}{0.95}\\

\hline

 & F nodes & V nodes & P nodes \\
 \hline
$\Delta$ (in m) & 400 & 200 & 100  \\
population & 54 & 50 & 200  \\
speed (in Km/h) & 0 & $[25-67]$ & $[2.5-6.5]$ \\
\hline
& \multicolumn{2} {|c|}{communication} & sensing \\
\hline
Memory policy & \multicolumn{2}{|c|}{SI/SD }&  SI/SD   \\
\hline
 \end{tabular}
\end{table}

\subsection{Information dynamics and energy savings}
\label{sec-information-policies}

The simplest policy that can be devised for reducing the overall number of transmissions (hence the overall energy consumption) is to increase the time between successive transmissions of each P node. To this end we consider increasing delays between transmissions; namely we analyze the average coverage for $T \in \{0.2, 10, 60, 600, \infty \}$ seconds. The value $T=\infty$ represents the extreme case when P nodes never transmit, i.e. they do not participate to the spreading of information; it represents a lower bound on the achievable performance. The analysis is carried out when information does not vary in time (the scenario is called \emph{static}) as well as when one randomly chosen information item is updated once every $d$ seconds. Some simulations based on this simple policy are already exploited in the static scenario in \cite{Corrente2012}.  

\begin{figure*}[t]
\centering
\includegraphics[width=70mm]{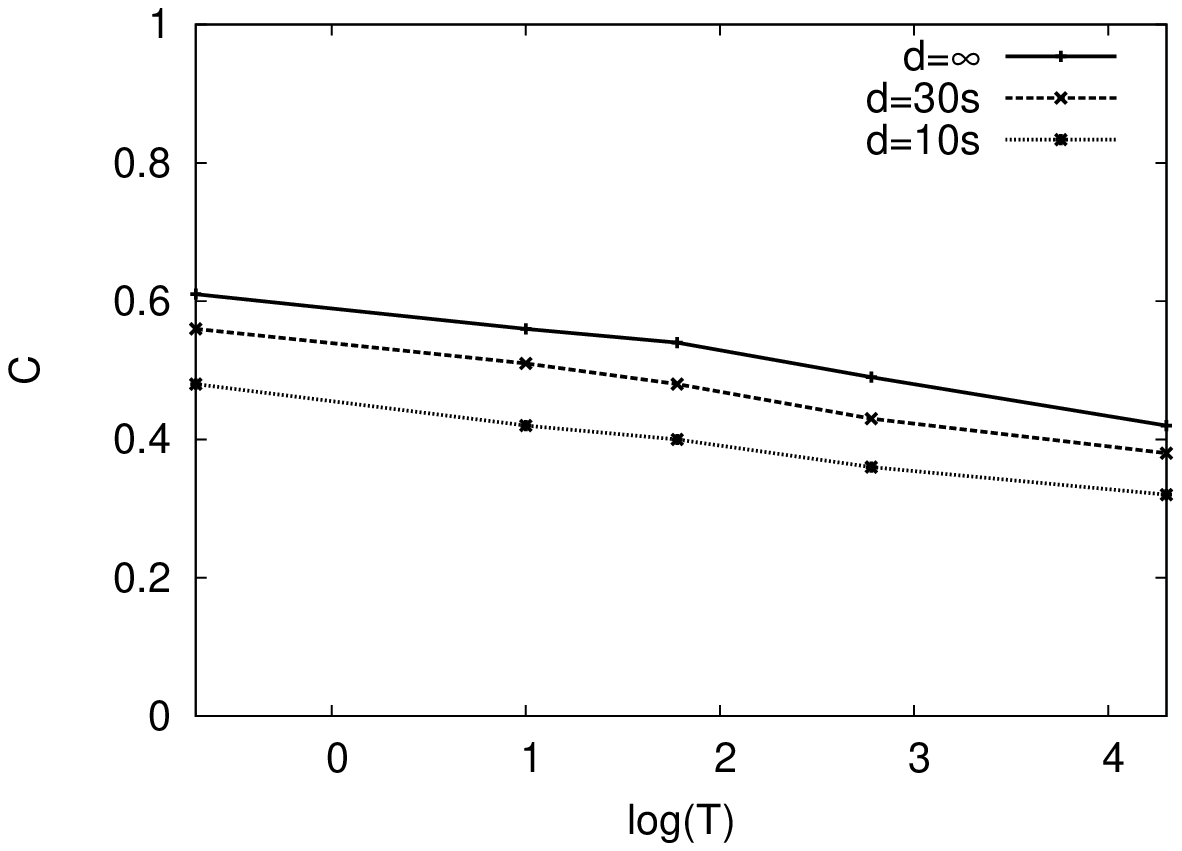}
\includegraphics[width=70mm]{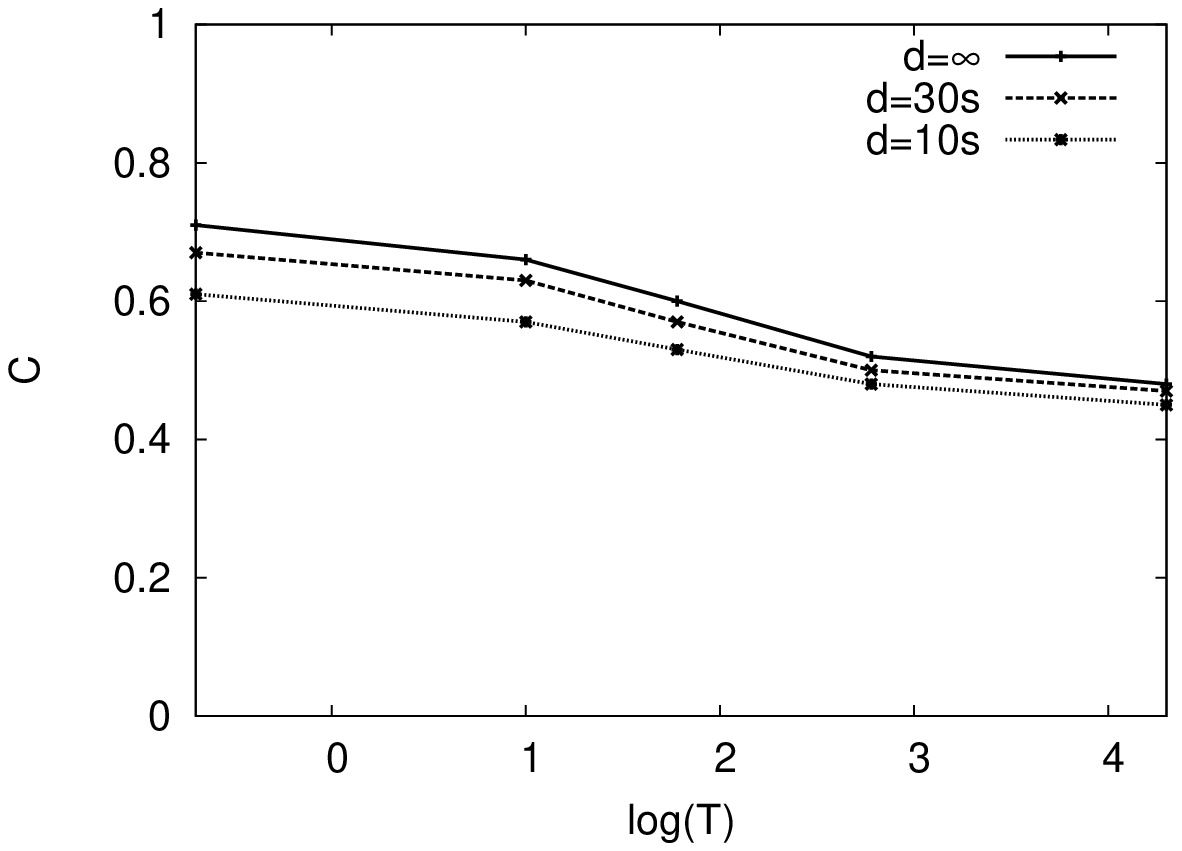}
\caption{Average coverage for P nodes (left) and V nodes (right) for increasing values of the transmission delay $T$.
\label{fig:update}}
\end{figure*}

Figure \ref{fig:update} shows the average coverage for different values of $d$ and for increasing delays between successive transmissions of pedestrians. Frequent information updates negatively impact on the average coverage of both pedestrians and vehicles. Pedestrians are able to sustain 1 update every 30 seconds with an acceptable performance reduction. Vehicles are much more resilient to information change and are able to efficiently cope with updates that occur once every 10 seconds. In Table \ref{tab:TransmPolicy} we have reported the overall number of transmissions for P nodes as a function of $T$ (first row). It can be noted that for $T=10$ seconds there is one order of magnitude less transmissions and for $T=60$ seconds the savings is two orders of magnitude.

We also devised more complex transmission policies. In particular, we leveraged the UDel Models capabilities of tracking the status of a node in the mobility traces; an attribute (called floor) is assigned to each node to classify its current position.
The floor attribute for a P node can be set to a code to represent any of the following cases:
\begin{itemize}
\item inside, the floor number a node is on;
\item outside simulated area;
\item walking outside;
\item driving in car;
\item in parking lot;
\item in subway station.
\end{itemize}
In practical deployments this context information can be obtained by the ever growing number of sensors being embedded in modern mobile phones, e.g., pressure, gyroscope, along with localization services provided by GPS, wifi and mobile networks.
This context information can be exploited to improve the information diffusion, taking into account the fact that it is very unlikely to get a radio contact between two terminals separated by a building element, e.g. one node walking outside and one sitting inside a building, or two nodes laying in separate floors of the same building.
According to this rationale it comes in useful to force a node to
spread the collected information as soon as its context changes, i.e. it is moving outside, inside or changing the floor, since it is very likely 
to contact novel nodes (unreachable in its previous context) increasing the probability to exchange fresh data.  
As a consequence we have devised the following context aware transmission policies:
\begin{itemize}
\item floor($w$): whenever a P node changes the value of the floor attribute the device wakes up for $w$ seconds and transmits once every $T=0.2$ seconds;
\item walk($w$): whenever a P node leaves or enters the walking outside state, e.g. whether it exits/enters a building or stops driving, it wakes up for $w$ seconds and transmits once every $T=0.2$ seconds;
\end{itemize}

\begin{figure*}[t]
\centering
\includegraphics[width=70mm]{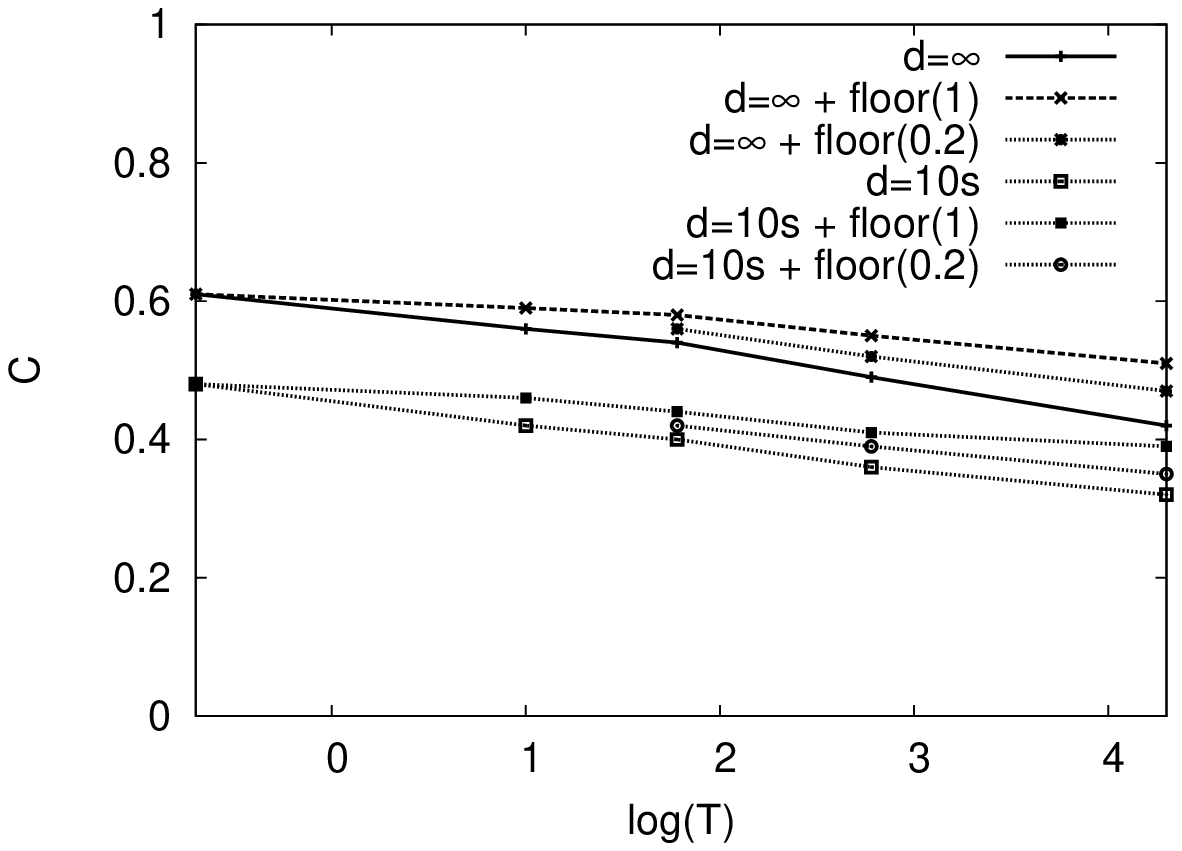}
\includegraphics[width=70mm]{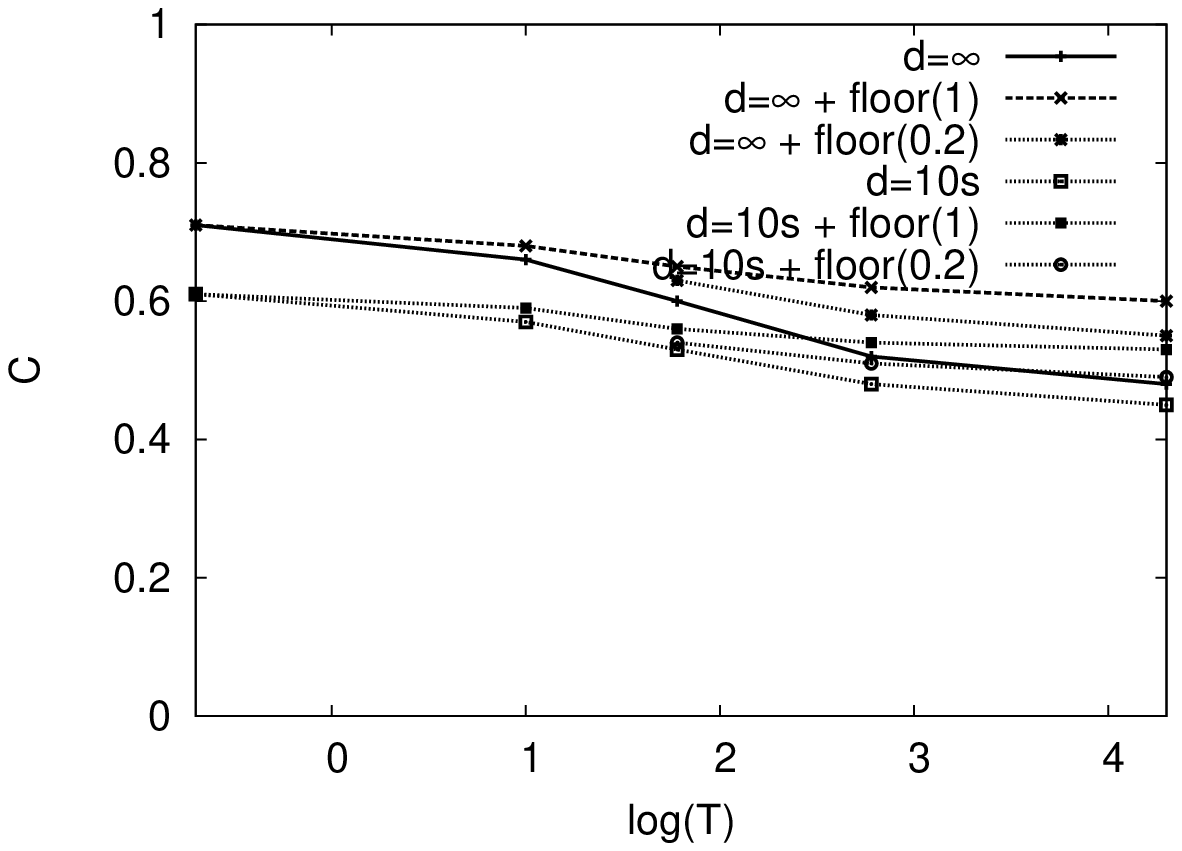}
\caption{Average coverage for P nodes (left) and V nodes (right) for increasing values of the transmission delay $T$ and for policies floor(1) and floor(0.2).
\label{fig:floor1}}
\end{figure*}

In the following experiment context awareness has been used in conjunction with previous approach based limiting the transmission opportunity. In Figure \ref{fig:floor1} we show the average coverage for P and V nodes when P nodes are allowed to transmit according to floor($w$), with $w=0.2$ s and $w=1$ s respectively, and at least once every $T$ s. 
It can be noted that, as we predict, there is an improvement of the average coverage for both P and V nodes for both static and dynamic information due to a better exploitation of P nodes mobility. 
In Table~\ref{tab:TransmPolicy} the overall number of P nodes transmissions (per simulated scenario) is shown along with the additional 
number of transmissions performed according to the context aware policies; it can be noted that walk(1) requires the lowest number of additional transmissions. On the other hand, Figure \ref{fig:floor1} shows that floor(1) achieves the best coverage. 

\begin{figure*}[t]
\centering
\includegraphics[width=70mm]{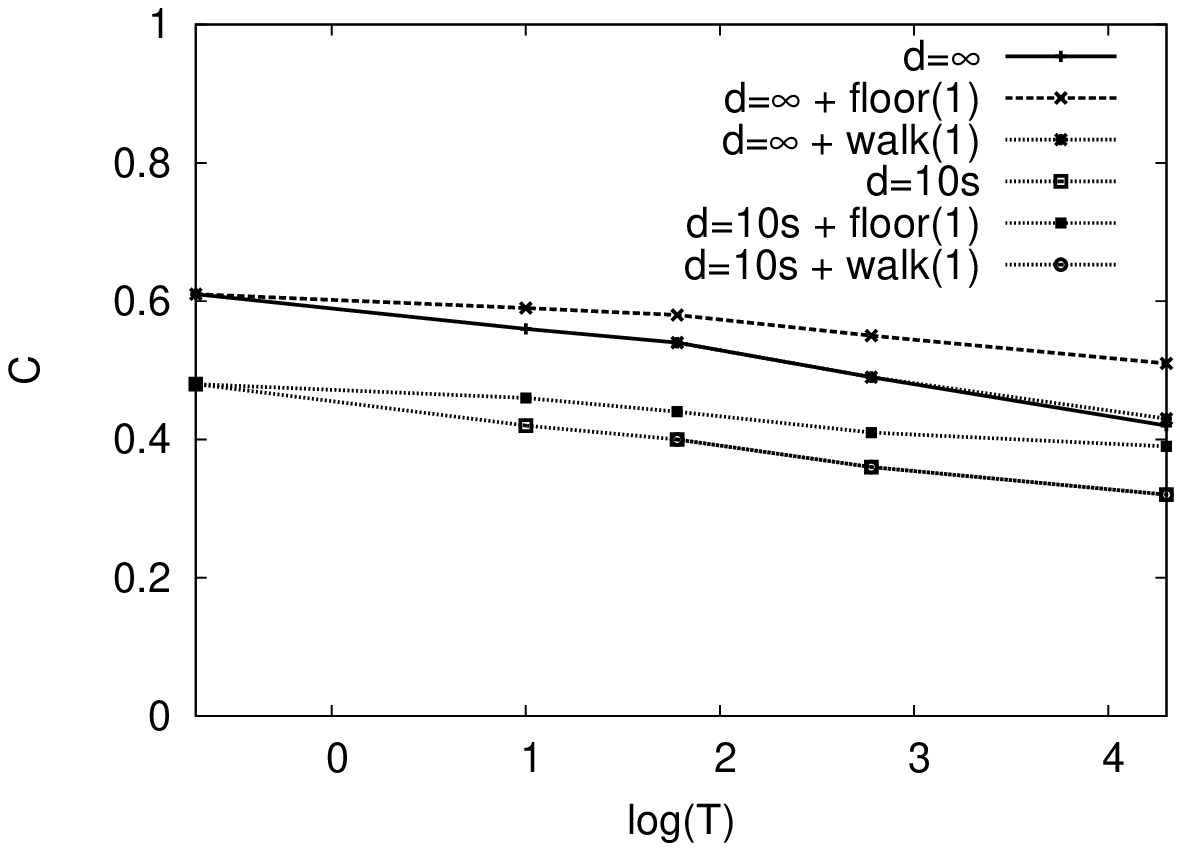}
\includegraphics[width=70mm]{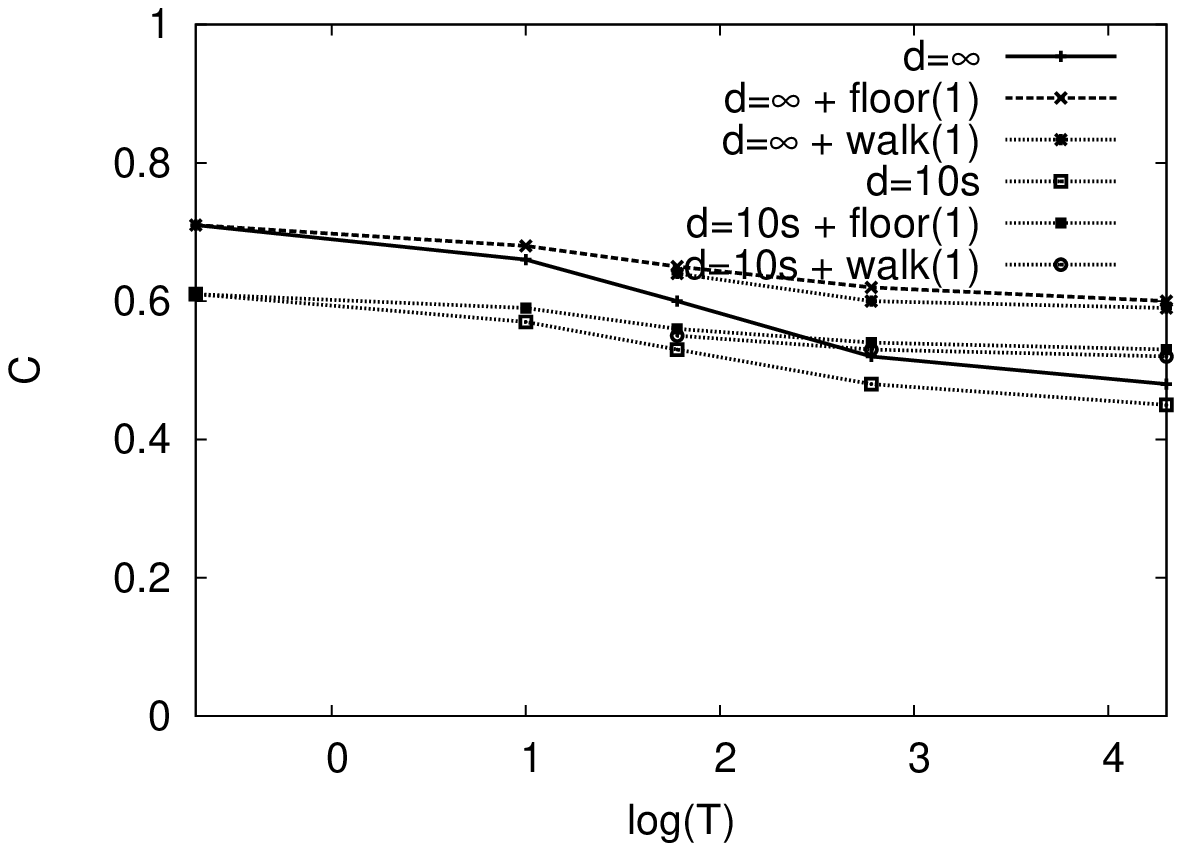}
\caption{Average coverage for P nodes (left) and V nodes (right) for increasing values of the transmission delay $T$ and for policies floor(1) and walk(1).
\label{fig:walk}}
\end{figure*}

Finally, Figure \ref{fig:walk} shows the comparison between floor(1) and walk(1) policies; although results in Table \ref{tab:TransmPolicy} clearly prove that the walk(1) policy can achieve a very small amount of transmissions for P nodes the gain in the average value of the coverage is negligible with respect to simply delaying successive transmissions. 

\begin{table*}[tb]
\centering
\caption{Overall number of transmissions for P nodes and overall additional transmissions for context aware policies.}
\label{tab:TransmPolicy}
\begin{tabular}{|c|c|c|c|c|c|}
\hline
 & \ T=0.2 & \ T=10 & T=60 & T=600 & T=$\infty$ \\
\hline
Reference scenario & 7000000 & 140000 & 23000 & 2300 & 0 \\
\hline
\multicolumn{6}{|c|}{Additional transmissions} \\
\hline
floor(1) & \multicolumn{5}{|c|}{6700} \\
\hline
floor(0.2) & \multicolumn{5}{|c|}{1340} \\
\hline
walk(1) & \multicolumn{5}{|c|}{260} \\
\hline
 \end{tabular}
\end{table*}

\subsection{Enhancing the building infrastructure}
\label{sec-elevators}

As we noted in the previous section, context awareness (implemented as additional transmission of P nodes based on the floor(w) and walk(w) policies) proved to be beneficial: there is an improvement of the average coverage for both P and V nodes for both static and dynamic information due to a better exploitation of P nodes mobility. These two policies rely on spontaneous nodes movements and context changing without the need of an extra infrastructure.

Nevertheless, we could improve the coverage of nodes by devising a simple solution based on elevators equipped with terminals able to relay data items. These terminals can be thought of as semi-fixed since their mobility is constrained to a vertical movement. We model the elevator mobility by means of two parameters: the average stopping time at a floor $W$ (60 seconds in our experiments) and the elevator speed (one floor per second). Every time an elevator moves it changes floor by choosing a destination uniformly at random among all floors of the building. Furthermore, their channel loss threshold was fixed to $-45$ dB.

In Table \ref{tab:PropChangeElevatorWait60Chloss45Delay60} we show the average coverage of P and V nodes for dynamic information ($d=10$) in the reference scenario. We also show the case of delayed transmissions ($T=60$) with and without the use of the floor(1) policy. 
It can be noted that E nodes allow the average coverage of both node types to increase in all cases. Indeed, when E nodes hold on at the lower floors of a building are able to collect information carried by V and P nodes passing close to the outer walls. These nodes hold data items that belong to the ROI of P nodes inside the building and vice-versa; in this case, E nodes act as trait d'union among P nodes inside the building and the surrounding external environment. Additionally, E nodes spread this information to all P nodes laying in upper floors as soon as they move up and down. In this case, E nodes act as trait d'union among P nodes laying at different floors of the same building.
The key role of E nodes is further highlighted in Table \ref{tab:PropChangeElevatorWait60Chloss45Delay60} where it is shown that short average waiting times at a floor translate into higher average coverage of P and V nodes.

\begin{table*}[tb]
\centering
\caption{Average coverage for P and V nodes with elevator based infrastructure.}
\label{tab:PropChangeElevatorWait60Chloss45Delay60}
\begin{tabular}{|c|c|c|c|c|c|c|}
\hline
 &  no E nodes             & $W$=180
&$W$=120&$W$=60&$W$=15& $W$=5\\
\hline
 & \multicolumn{6}{|c|}{P nodes} \\
\hline
delayed   & 0.40 &  0.50      & 0.51 & 0.55 &0.61 & 0.63\\
\hline
delayed + floor(1)      & 0.44 &    0.54   & 0.55 & 0.58 &0.63& 0.66\\
\hline
reference scenario        &  0.48 &   0.59&  0.60 &0.64& 0.70&0.73 \\
\hline
 & \multicolumn{6}{|c|}{V nodes} \\
\hline
delayed  &  0.53&    0.64      &0.64& 0.66 &0.67& 0.68\\
\hline
delayed + floor(1)       & 0.56 &    0.67 &0.67& 0.68 &0.70& 0.71\\
\hline
reference scenario        &  0.61 &   0.72 &0.72& 0.75&0.77 & 0.78\\
\hline
\end{tabular}
\end{table*}

\subsection{ROI dynamics}
\label{sec:roi}
In all previous analysis we assume that every node is interested in retrieving
the information located around its own position. As a consequence, the ROI has
been linked with node position. 
In the following,  this constraint will be removed letting the nodes pick up a random ROI according to several models.
Table~\ref{tab:ROIdyn} compares the coverage obtained by P and V nodes in several scenarios varying the ROI model, the caching policies and  transmission frequency.
In particular, the results obtained in the reference scenario with a linked ROI are compared with those worked out when nodes let their ROIs floating according to the following model: every 10 minutes a node selects the ROI corresponding to its instantaneous position, then the node moves according to
the mobility model without updating the ROI location. 
The first two sets of experiments in Table~\ref{tab:ROIdyn} investigate the effects of the selective insert (SI) policy in the static scenario ($d=\infty$). In particular, we investigate the effect of using SI
when reading a piece of information from the environment, i.e. associated to
the node position. It can be noted that using SI when the ROI location does 
not coincide with the node position can be very critical, because it represents an egoistic behavior where data items that are not useful (in terms of ROI) for a node are not acquired and shared in the opportunistic network. This conjecture is confirmed by the simulations, where we notice a slight improvements in terms of coverage for the floating ROI in the case not using SI for environment sensing. As a consequence, SI is not used in all the remaining results of Table~\ref{tab:ROIdyn}, where we analyze the performance when the information is updated with $d=10$ s, delayed transmissions ($T=60 s$) and when elevators are used to help spreading the data. 
The simulation results show that the proposed approaches are effective also for the floating ROIs. In particular, for P nodes we report very limited differences with respect to the case of linked ROIs. This is due to the fact that P nodes mobility within 10 minute is limited, thus the linked and floating ROI models are quite similar in practice. On the other hand, V nodes significantly improve their coverage in the case of floating ROI. Indeed, fixing the ROI for 10 minutes, makes
it easier for faster V nodes to collect the items of interest because these are kept constant for a certain period.

\begin{table*}[tb]
\centering
\caption{Coverage of P and V nodes for linked ROI (l-ROI) 
and floating ROI (f-ROI) for different policies.}
\label{tab:ROIdyn}
\begin{tabular}{|p{4.8cm}|cc|cc|cc|}
\hline
 & \multicolumn{2}{|c|}{scenario} & \multicolumn{2}{|c|}{C for P nodes} &  \multicolumn{2}{|c|}{C for V nodes} \\
  & $N_P$ & $N_V$ & l-ROI & f-ROI & l-ROI & f-ROI \\
\hline
\multirow{2}{*}{\vbox{$d=\infty$, with SI}} 
& 200 & 50 & 0.62 & 0.62 & 0.71 & 0.73 \\
& 500 & 0 & 0.62 & 0.62 & - & - \\
\hline
\multirow{2}{*}{\vbox{$d=\infty$, no SI}}
& 200 & 50     & 0.62 & 0.66  & 0.71 & 0.81 \\
& 500 & 0     & 0.62 & 0.66     & -       & -      \\
\hline
\multirow{2}{*}{\vbox{$d=10$ s, no SI}} 
& 200 & 50  & 0.49 & 0.47 & 0.62 & 0.72 \\
& 500 & 0 & 0.51 & 0.48 & - & - \\
\hline
\multirow{2}{*}{\vbox{$d=\infty$, no SI, delayed }} 
& 200 & 50  & 0.54 & 0.59 & 0.62 & 0.73 \\
& 500 & 0  & 0.53 & 0.57 & - & - \\
\hline
\multirow{2}{*}{\vbox{$d=10$ s, no SI, delayed }} 
& 200 & 50  & 0.41 & 0.40 & 0.53 & 0.65 \\
& 500 & 0  & 0.40 & 0.39 & - & - \\
\hline
\multirow{2}{*}{\vbox{$d=\infty$, no SI, E nodes}} 
& 200 & 50  & 0.80 & 0.84 & 0.86 & 0.91 \\
& 500 & 0  & 0.79 & 0.81 & - & - \\
\hline
\multirow{2}{*}{\vbox{$d=10$ s, no SI, E nodes}} 
& 200 & 50  & 0.64 & 0.60 & 0.74 & 0.81 \\
& 500 & 0 & 0.65 & 0.61 & - & - \\
\hline 
\end{tabular}
\end{table*}

Finally, a static ROI model has been simulated, where all nodes randomly select one out of
$N_{ROI}$ static ROIs, with a given time frequency. In Table~\ref{tab:in-out} the coverage 
obtained in the limit case of $N_{ROI}=1$, i.e. same area of interest for all nodes, is shown.
In particular, we show the sensitiveness of $C$ with respect to the position of the ROI inside 
and outside the buildings, respectively. It can be noted that P nodes are more effective in collecting  information inside buildings, where the density of P nodes is higher.
On the contrary, the V nodes coverage improves when the area of interest is located outside.

\begin{table}
\centering
\caption{Coverage of P and V nodes with a single static ROI placed
outside or inside buildings in the case $N_P=200$, $N_V=50$, time 7-10 (reference scenario with $d=\infty$ without and with E nodes).}
\label{tab:in-out}
\begin{tabular}{|c|cc|cc|}
\hline
  & \multicolumn{2}{|c|}{C for P nodes} &  \multicolumn{2}{|c|}{C for V nodes} \\
   & out & in & out & in \\
\hline
reference & 0.55 & 0.70 & 0.89 & 0.55  \\
with E nodes & 0.65 & 0.87 & 0.92 & 0.86 \\
\hline

\end{tabular}
\end{table}

\begin{figure*}[tb]
\centering
\includegraphics[width=70mm ]{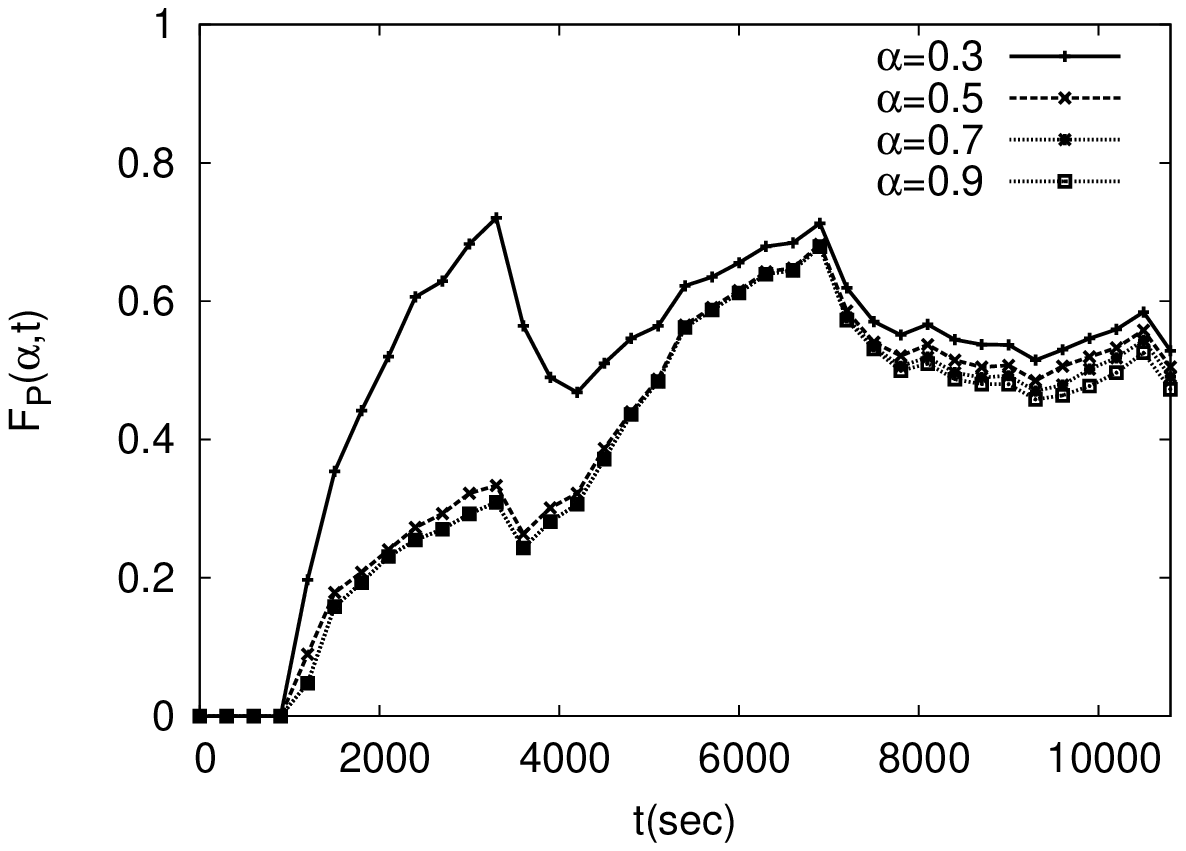}
\includegraphics[width=70mm]{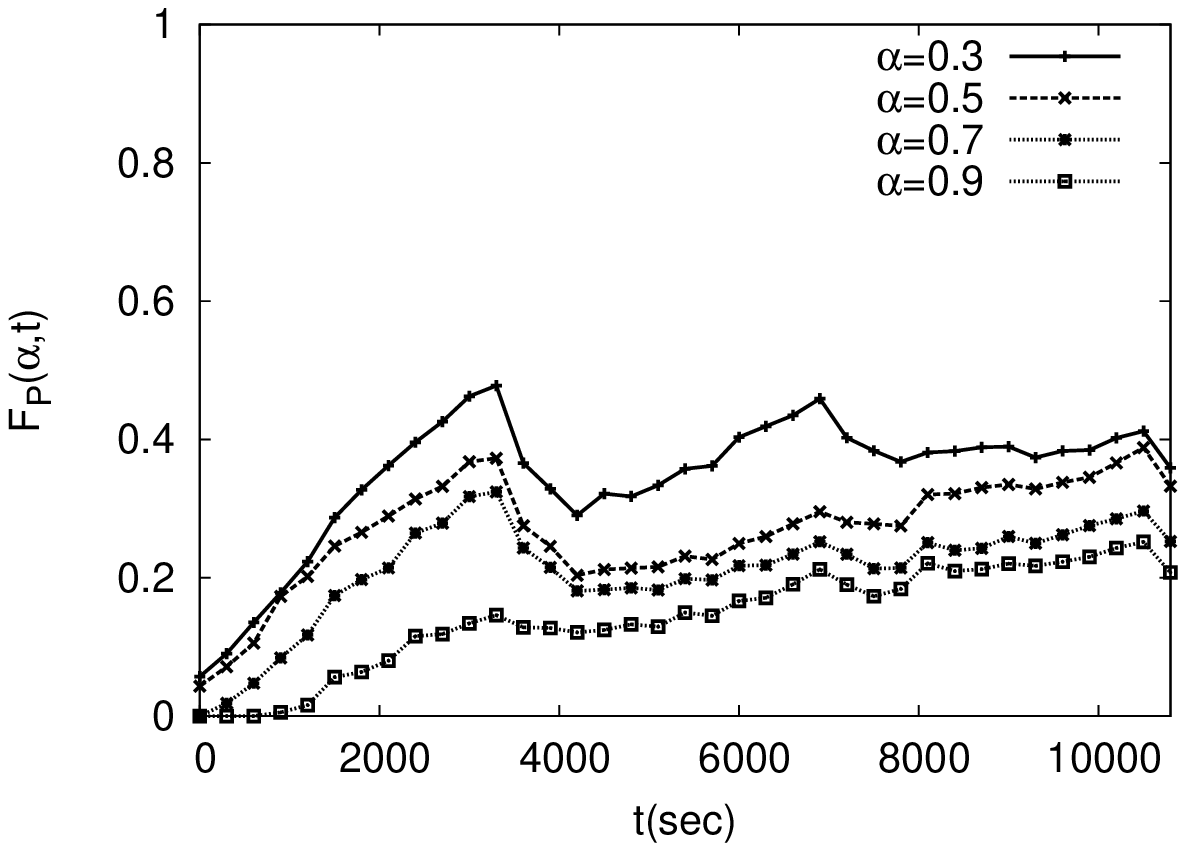} \\
\includegraphics[width=70mm]{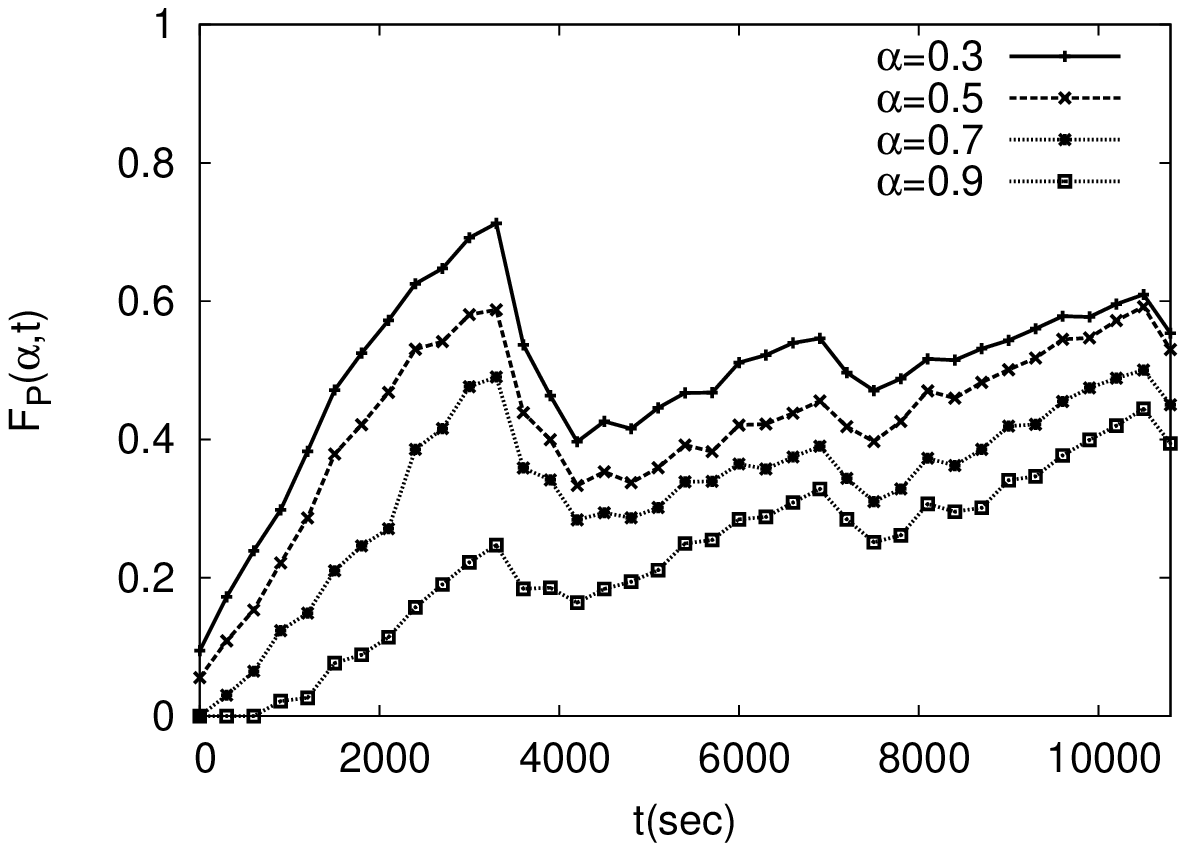}
\includegraphics[width=70mm]{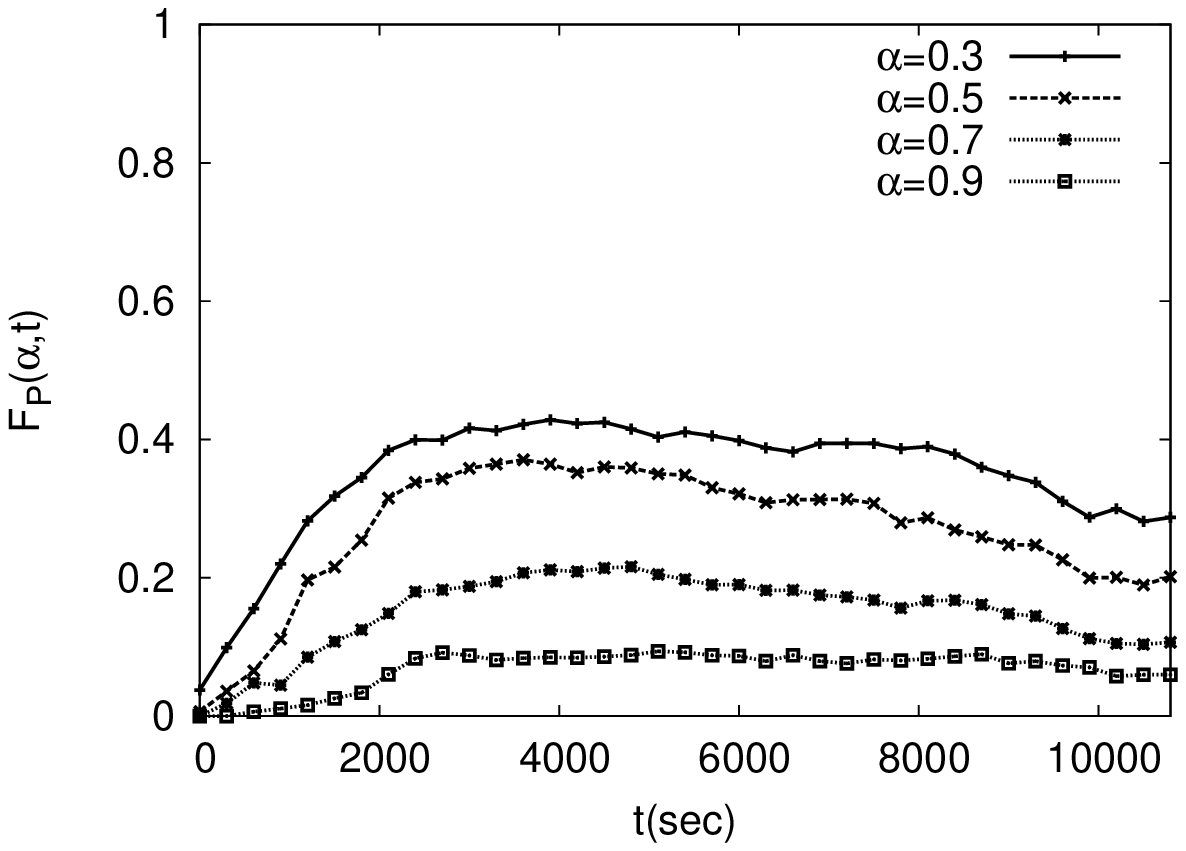}

\caption{Temporal behavior of $C$ for P nodes in the reference scenario with $d=\infty$ when $N_{ROI}=1$ with 1 hour synchronous update  (top left), $N_{ROI}=4$ with 1 hour synchronous update (top right), $N_{ROI}=4$ with 1 hour synchronous update and $E$ nodes (bottom left), $N_{ROI}=N$ with 1 hour asynchronous update (bottom right).
\label{fig:4ROI}}
\end{figure*}

In presence of ROI dynamics it is very interesting to analyze the temporal behavior of the coverage. To this end, we define the performance index $F_X(\alpha,t)$ that represents the
fraction of nodes of class $X$ having $C\geq\alpha$ at time $t$. In Figure~\ref{fig:4ROI} we show $F_P(\alpha, t)$, i.e. the percentage of P nodes  with $C \geq\alpha = 0.3, 0.5, 0.7, 0.9$ as a function of $t$. 
This plots allow one to appreciate the distribution of the values of $C$ of the P nodes.
Figure~\ref{fig:4ROI} refers to the the reference scenario with static information ($d=\infty$). The top-left figure is worked out in the case of
a single ROI being updated randomly every 1 hour, synchronously by all nodes.
The synchronous update is used as a limit case where all nodes change their interests
at the same time and allow us to appreciate the worst case delays in the propagation of the information. As a consequence, in the plot the three transient behaviors due to the synchronous 
updates turns to be evident. The top-right figure refers to the same synchronous ROI update
experiment when increasing the number of ROIs to $N_{ROI}=4$; it can be observed that nodes
coverage reduces when $N_{ROI}$ increases. The positive effects yielded by the exploitation 
of $E$ nodes is shown in the third figure (bottom-left), where the coverage distribution is 
highly improved in the case $N_{ROI}=4$.
Finally, in the bottom-right part of Figure~\ref{fig:4ROI} we show the coverage distribution when
all nodes randomly updates their ROIs every hour in an asynchronous fashion. In this latter case,
every node picks up a random ROI and keeps it fixed for a random period uniformly distributed between 50 and 70 minutes.

%% file: CONCLUSION.tex
\section{Conclusions}
\label{sec:conclusions}

We analyzed the performance of the system by defining the node coverage, i.e., the percentage of
the ROI covered by the updated items stored in the memory of each node. 
We carried on a simulative study to investigate the detailed mobility and radio propagation traces generated by the UDelModels tools to estimate this performance index for all nodes types. The main findings are:
\begin{itemize}
\item lowering the transmission rate of devices operated by pedestrians is an effective strategy for energy savings without significant coverage reduction even under information dynamics;
\item adding context awareness strategies further improves the results;
\item coverage of all nodes types can be increased by providing buildings with an infrastructure of relaying nodes mounted on elevators;
\item coverage of nodes increases when nodes share the same (few) ROIs that remain fixed for longer periods of time.
\end{itemize}

The current work may be extended in several ways: we are considering the impact of coding techniques \cite{LT} on the coverage of nodes as well as the use of compressive sensing \cite{CS1,CS2} concepts for sparsely defined information items. Finally, we are currently exploring the research area on spatio-temporal databases \cite{SPATIO-TEMPORAL} to import ideas and techniques to improve our work.